# Conceptual Opto-Mechanical design of SHARP a near-infrared multi-mode spectrograph conceived for the next-generation telescopes[1]

**Hossein Mahmoodzadeh[a,b*], Paolo Saracco[a], Paolo Conconi[a], Bortolino Saggin[b], Diego Scaccabarozzi[b], Ivan Di Antonio[b], Marco Riva[a], Emilio Molinari[a], Carmelo Arcidiacono[d], Ilaria Arosio[b], Enrico Cascone[e], Vincenzo Cianniello[e], Vincenzo De Caprio[e], Gianluca Di Rico[c], Benedetta Di Francesco[c], Christian Eredia[e], Paolo Franzetti[f], Marco Fumana[f], Davide Greggio[d], Elisa Portaluri[c] and Marcello Scalera[b]**

[a]INAF - Osservatorio Astronomico di Brera, Milano, Italy

[b]Politecnico di Milano, Mechanical Engineering, Milano, Italy

[c]INAF - Osservatorio Astronomico d'Abruzzo, Teramo, Italy

[d]INAF - Osservatorio Astronomico di Padova, Padova, Italy

[e]INAF - Osservatorio Astronomico di Capodimonte, Napoli, Italy

[f]INAF - IASF, Milano, Italy

**Abstract**. The next generation of Extremely Large Telescopes (ELTs), with their wide apertures and advanced Multi-Conjugate Adaptive Optics (MCAO) systems, will provide unprecedented sharp and deep observations, even surpassing the capabilities of James Webb Space Telescope (JWST). SHARP, a near-infrared (0.95-2.45 μm) spectrograph, is designed to optimally exploit the collecting area and angular resolution of these forthcoming ELTs, and specifically optimized for the MCAO unit MORFEO at the ELT. SHARP includes two main units: NEXUS, a Multi-Object Spectrograph (MOS), and VESPER, a multi-Integral Field Unit. This paper outlines the opto-mechanical design of SHARP based on the scientific requirements of the project. The optical design is engineered to meet project specifications, featuring a compact mechanical structure that minimizes the required cryogenic power while ensuring ease of access for maintenance and straightforward assembly procedures.

*****First Author**, E-mail: Hossein.Mahmoodzadeh@Polimi.it

## 1 Introduction

The upcoming generation of ground-based and space-based high angular resolution telescopes, such as the Adaptive Optics (AO)-assisted Extremely Large Telescopes (ELTs) and the Habitable World Observatory, will deliver sharper and deeper capabilities than the James Web Space Telescope (JWST). In particular, the European Southern Observatory's Extremely Large Telescope is poised to deliver unprecedented observational data, offering imaging with six times greater sharpness than JWST. This remarkable capability is attributed to the synergy between the world's largest aperture and the cutting-edge Multi-Conjugate Adaptive Optics (MCAO) system,

---

MORFEO[1,2], which is designed to correct atmospheric turbulence not only along the telescope's line of sight but also uniformly over a ~1.8'x1.8' field.

Table 1: SHARP main requirements resulting from the scientific drivers.

| | |
|---|---|
| Wavelength limit | 2.45 μm |
| Spectral ranges (simultaneous) | 0.95-2.45 μm |
| Angular resolution | ~ 30 mas |
| AO-corrected FoV | ~ 1x1 arcmin |
| Observing modes | MOS + multi-IFU |
| Spectral resolution | R>1000 |

MORFEO will initially be integrated with ELT's first-light instrument, MICADO[3,4], which is a near-infrared camera designed for imaging and long-slit spectroscopy, capable of producing images up to six times sharper than those of JWST. MORFEO is also designed to support a second instrument in the future, which is expected to be a spectrograph capable of fully exploiting the large aperture of ELT and the large, corrected field by MORFEO providing spectroscopic follow up to MICADO observations.

This paper presents a study for the conceptual opto-mechanical design of SHARP, a multi-mode Near-IR spectrograph designed to leverage the capabilities of upcoming MCAO systems on ELTs or any other high-angular resolution telescope. The development of the SHARP conceptual design is aimed at responding to the expected calls for new instrumentation for these telescopes. Its primary target is MORFEO at the ELT, since the latter is the telescope with the largest aperture (about 39 m diameter), while the former provides the widest AO-corrected field. Therefore, the conceptual study of SHARP[5] is currently being developed to maximize the capabilities offered by MORFEO at the ELT. SHARP offers an angular resolution and a range of spectral resolutions that surpass those of NIRSpec[6], the optical and near-IR spectrograph on JWST[7]; These features combined with a multiplex capability allow SHARP to fully exploit MORFEO's large corrected field of view.

SHARP is composed of two primary units: NEXUS, a Multi-Object Spectrograph (MOS), and VESPER, a multi-Integral Field Unit (mIFU). The scientific requirements for SHARP are summarized in Table 1, while a discussion on these scientific drivers can be found in Saracco et al[8]. Briefly, the requirements arise from the need to be able to detect particular atomic emission (e.g., Lyα[1216], HeII[1640], OII[3727], Hα[6563]) and absorption lines (e.g., Mgb[5120], CaH&K[3950], D[4000], MgII[2800]) in the spectrum of very distant galaxies, (i.e. at high redshift z), and to spatially resolve sizes comparable to those of giant molecular gas clouds (~150-200 pc) over the wide redshift range (see Fig. 2 in Saracco et al.[8]). The atomic emission and absorption lines above fall in the near infrared, specifically at $\lambda_{obs}$>1.8 μm for galaxies at z>2 (where $\lambda_{obs}=\lambda\times(1+z)$). Therefore, to study the properties of galaxies at z>2 that populate the Universe in the first quarter of its life, the primary requirement is to push the near-IR limit to $\lambda_{lim} \simeq$ 2.45 μm, where sky transmission remains high and sky emission can be still efficiently removed. Furthermore, an angular resolution of about 30 mas is needed to resolve the typical sizes of the giant molecular clouds over the whole cosmic time. Finally, the detection and study of the needed spectral features require a resolution R = $\lambda/\Delta\lambda$ > 1000.

## 2    Concept and optical paths of SHARP

Figure 1 illustrates the key components along the optical path from the port of MORFEO to SHARP and the optical path of the light within SHARP and through its key components.

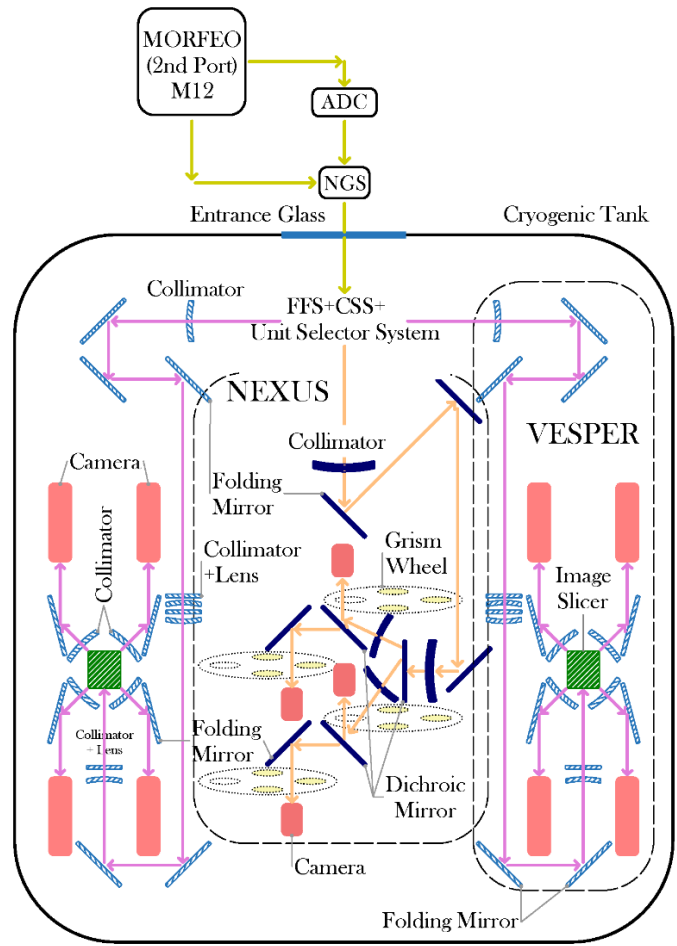

**Figure 1:** components and optical path from the 2nd port of MORFEO to SHARP and optical path of SHARP and its key components. SHARP is fed by an ADC and a NGS unit.

SHARP, is fed by an Atmospheric Dispersion Corrector (ADC) and a Natural Guide Stars (NGS) unit. Briefly, the light from MORFEO enters SHARP after passing through the ADC. The ADC plays a crucial role by correcting for atmospheric dispersion, ensuring that SHARP can simultaneously collect spectra across the entire wavelength range, regardless of the zenith distance or rotation angle.

We notice that for IFU observations atmospheric dispersion correction can be obtained via post-processing. Therefore, a removable ADC from the beam in the case of VESPER observations could enhance the sensitivity optimizing the SHARP performance. The concept study of the ADC is beyond the scope of this paper; we leave its description to a forthcoming paper.

The NGS unit, which includes wavefront sensors monitoring up to three NGSs, provides real-time feedback to MORFEO. This information is used in conjunction with six laser guide stars (LGS) to actively compensate for atmospheric turbulence across the field of view. This field, at the second port, spans a diameter of 160 arcseconds and encompasses the full field of view of SHARP's spectrographs. Notably, the NGS unit used by SHARP shares its design with the unit feeding the MICADO instrument, albeit without the SCAO (Single-Conjugate Adaptive Optics) module, which is not required for SHARP.

Just below SHARP's entrance window is the Unit Selector System (USS), the component that directs the incoming light to either NEXUS or VESPER. according to the observational requirements. The USS hosts the two mechanisms that feed NEXUS and VESPER, namely the Configurable Slit System (CSS), a multi-slit system, and the Field Selector System (FSS), a system to deploy the probes, respectively. A description of these two systems is given on Sec. 4.

Notably, SHARP's optical design is entirely free of aspheric surfaces, a choice that simplifies manufacturing and alignment while maintaining the optical performance required for its scientific goals. The optical design and expected performance will be detailed in a dedicated paper.

SHARP hosts 12 cameras., Four cameras are used by NEXUS (MOS) to simultaneously cover the whole wavelength range 0.95-2.45µm (see Sec. 2.1); eight cameras feed VESPER (mIFU) that, in the current configuration, is composed of two modules (four cameras per module). The two modules support 12 probes (6 per module) (see Sec. 2.2).

## 2.1 NEXUS: Multi-Object Spectrograph

NEXUS, the Multi-Object Spectrograph, is designed to operate over an AO-corrected field of approximately 1.2'×1.2' with a pixel scale of 35 mas, as illustrated in Figure 2.

The optical design of NEXUS, depicted in Figure 3, begins at the focal plane of MORFEO, where the CSS is located. The CSS mechanisms can deploy up to 30 slits, each measuring approximately 2.4" (see Sec. 4.7). The light is then folded by three folding mirrors, a design choice that compacts the instrument while maintaining optical precision. The light coming out of the CSS is split into four distinct wavelength bands, approximately [0.95-1.15] μm, [1.15-1.45] μm, [1.45-1.9] μm, and [1.9-2.45] μm, using three dichroic filters (Figure 1). Each band is fed by a dedicated camera hosting a grism wheel, with each camera optimized for its specific wavelength range. This allows NEXUS to simultaneously cover the whole wavelength range, without having to switch from one configuration to another significantly enhancing the instrument and telescope efficiencies. For comparison, EMIR[9-11] (Espectrógrafo Multiobjeto Infrarrojo), the spectrograph on the Gran Telescope Canarias (GTC), requires switching among different optical elements for observations at different wavelength ranges. Similarly, NIRSpec[12,13] on the JWST covering a broader wavelength range of (0.6 - 5.3 µm) employs different configurations to operate in different frequency bands. MOSFIRE[14] (Multi-Object Spectrometer For Infra-Red Exploration) at the 10 m Keck provides near-infrared (0.97 to 2.45 µm) multi-object spectroscopy utilizing a single diffraction grating that can be set at two fixed angles, with order-sorting filters to cover the K, H, J, or Y bands.

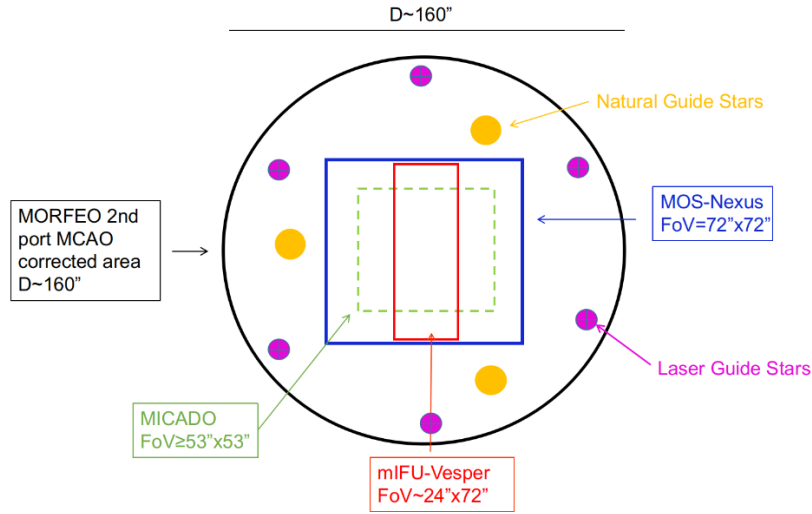

**Figure 2:** The Field of View (FoV) of NEXUS (blue square, 1.2'×1.2') and the area probed by the IFSs of VESPER (red rectangle, ~24"×70") are shown on the MORFEO 2nd port corrected area (black circle, diameter D~160"). For comparison, the FoV of MICADO (green dashed square) is also shown. The purple filled circles represent the wavefront sensors for the 6 Laser Guide Stars used by MORFEO, while the yellow circles are those for the 3 NGSs.

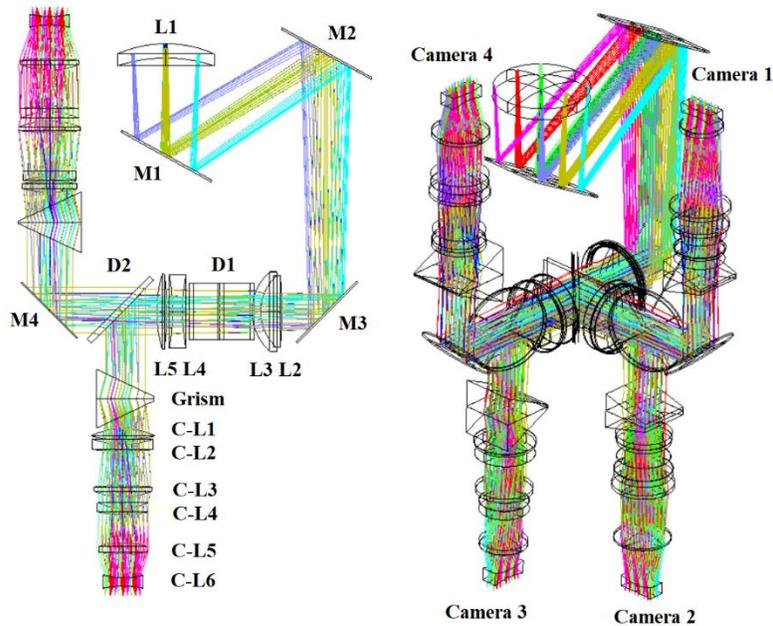

**Figure 3:** The optical design of NEXUS, including the optical path from the focal plane of MORFEO down to the individual cameras.

The optical elements of the NEXUS are described in Table 2. Each camera of NEXUS is powered by 4 2kx2k detectors, summing up to 8000 pixels in the dispersion direction. This number of pixels is necessary to collect the whole spectrum for all the sources independently of their position in the field and of the selected spectral resolution. The cameras feature a grism wheel housing three grisms each with $R=\lambda/\Delta\lambda=300, 2000, 6000$ for a reference slit width of 0.2 arcsec(An accurate subtraction of OH lines require a spectral resolution higher than R~3000. However, not all the science cases need an accurate subtraction of OH lines. Some of them benefit from a higher S/N (lower spectral resolution) rather than from an accurate OH lines subtraction. For example, a resolution R~6000 is best suited for studies of atomic line profiles (second and third order moments) and for measurements of narrow lines, since it provides good accuracy in the subtraction of OH lines. In contrast, other studies focused on the mean properties of the stellar population and on the physical properties of galaxies (e.g., stellar velocity dispersion, metallicity, abundances) require a lower spectral resolution (R~2000) being less influenced by the presence of OH lines while strongly dependent on the signal-to-noise ratio. The lowest resolution, R~300, is used to maximize the signal-to-noise ratio of the continuum of extremely faint objects. It is intended to derive the shape and detect breaks in the spectrum of the faintest objects, for which fine subtraction of the OH lines is not relevant). To understand how many pixels and hence how many detectors are needed to collect a spectrum at specific resolutions, we consider the example of an observation at R~6000 at 2.1 µm, where this configuration results in ~3200 pixels. If the observed source is in the center of the NEXUS field two detectors of 2k pixels would be sufficient to collect the whole spectrum. However, to collect a complete spectrum also for sources close to the borders of the field, e.g. displaced ±35" by the center, we should consider ±1600 pixels more, summing up to 6400 pixels. Therefore 4 detectors 2k are needed. For a point source the spectral resolution is R~17000 for the highest resolution grism (The slit width that can be seen by a point source (whose size is virtually null) is given by the diffraction limit of the telescope, Dl=0.012 arcsec (12 mas) for ELT at 2.1 µm. Therefore, considering a resolving power R(0.2″)≈6000 (for a slit width of 0.2″) and scaling according to the slit width (sw), we can write the resolving power as R(sw)=(0.2″/sw)×R(0.2″). Using this, for a diffraction-limited slit width of 12 mas, we obtain R(12 mas)=(0.2″/0.012″)×6000≈100000. This is the theoretical spectral resolution expected for a

**Table 2:** Overview of the NEXUS optical components, including key parameters and design characteristics

| | R1 [mm] | R2 [mm] | Thickness [mm] | Material |
|---|---|---|---|---|
| **The lens from focal panel up to Cameras** | | | | |
| L1 | 419.05 | 531.185 | 50 | SILICA |
| L2 | 1150.9 | 9152.1 | 30 | CAF2 |
| L3 | 169.5 | 188.6 | 30 | CLEARTRAN |
| L4 | 725.4 | 512.04 | 30 | SILICA |
| L5 | 953.1 | 508.2 | 30 | CAF2 |
| L6 | 725.4 | 512.04 | 30 | SILICA |
| L7 | 953.1 | 508.2 | 30 | CAF2 |
| **Camera 1** | | | | |
| C-L1 | 450.1 | 2268.1 | 40 | CAF2 |
| C-L2 | 313.3 | 575 | 30 | SILICA |
| C-L3 | 1072.6 | 975.1 | 25 | SAPPHIRE |
| C-L4 | 639.9 | 4712.1 | 25 | CLEARTRAN |
| C-L5 | 248.6 | 724.1 | 25 | SAPPHIRE |
| C-L6 | 344.4 | 593.6 | 25 | SAPPHIRE |
| **Camera 2** | | | | |
| C-L1 | 290.3 | 2486.1 | 40 | CAF2 |
| C-L2 | 412.4 | 267.2 | 30 | SILICA |
| C-L3 | 1414.9 | 4847.5 | 25 | SILICA |
| C-L4 | 927.3 | 1401.7 | 25 | CLEARTRAN |
| C-L5 | 288.7 | 584.3 | 35 | CAF2 |
| C-L6 | 207.1 | 677.1 | 25 | SILICA |
| **Camera 3** | | | | |
| C-L1 | 268.7 | 410.7 | 40 | CAF2 |
| C-L2 | 304.3 | 3017,1 | 20 | S-TIM28 |
| C-L3 | 979.1 | 262.4 | 40 | N-PSK53 |
| C-L4 | 192.3 | 520.5 | 20 | S-TIM28 |
| C-L5 | 279.1 | 458.5 | 40 | N-PSK53 |
| C-L6 | 222.4 | 299.9 | 35 | S-TIM28 |
| **Camera 4** | | | | |
| C-L1 | 358,3 | 425.1 | 40 | CAF2 |
| C-L2 | 313.2 | 5525.8 | 20 | S-TIM28 |
| C-L3 | 981.6 | 276.8 | 40 | N-PSK53 |
| C-L4 | 224.6 | 484.1 | 20 | S-TIM28 |
| C-L5 | 291 | 721.1 | 40 | N-PSK53 |
| C-L6 | 222.4 | 299.9 | 35 | S-TIM28 |

point source in case of a diffraction limit of 12 mas. However, the pixel size of NEXUS is 35 mas. This means that a diffraction-limited image (12 mas) falls onto approximately 3 pixels (35 mas/12 mas≈2.9). Because the instrument's effective resolution is limited by the pixel size when it's larger than the diffraction limit, the actual resolving power becomes

R≈100000/(35 mas/12 mas)≈100000/2.9≈34,483. The Nyquist criterion states that to accurately sample a signal at least two samples per resolution element are needed. Therefore considering two pixels the expected resolving power is ~17000). This design ensures that NEXUS can deliver high-resolution spectra across a wide range of scientific applications. The expected throughput in K is ~90%, neglecting the grism.

The schematic view of NEXUS is also depicted in Figure 1.

## 2.2 VESPER: Multi-Integral Field Unit (mIFU)

VESPER, the second main subsystem of SHARP, is an mIFU designed to perform spatially-resolved spectroscopy over the wavelength range of 1.2 to 2.4 μm. VESPER is specifically engineered to capture spectral information from multiple regions within a single field of view, making it ideal for studies requiring detailed mapping of complex astrophysical phenomena, such as the kinematics of galaxies or the ionization within star-forming regions.

The optical layout of VESPER, as shown in Figure 4, originates at the focal plane of MORFEO, where the FSS is situated (together with the CSS of NEXUS). Then, light enters VESPER via a series of folding mirrors and subsequently goes through some optical elements before reaching the image slicer placed onto the VESPER focal plane (see Figure 1). The summary of the VESPER optical elements is provided in Table 3.

VESPER is a modular system. Each module is composed of 6 probes called Field Selectors (FS) having a field of view of approximately 1.7"×1.5" each.

The 6 FSs of a VESPER module are arranged in a Cartesian xy grid, with a separation of 0.33" along the x-axis (see Figure 5). The FSs can be deployed over a range of ~70" along the y-axis. The movement is compensated by two moving mirrors to keep constant the optical path of all the FSs so that they form on the focal plane a rectangular image of size Y×X=1.5" × [(1.7"×6)＋0.33"×5)]~1.5"×12", as if they were aligned. The gaps between the FSs are masked at the focal plane to eliminate background contributions from scattered light. The slicer placed on the focal plane of VESPER is composed of four sets of 72 micro-mirrors each. Each set samples a stripe equal to 1/4 of the image that is 0.375"×12" (see Figure 5), and sends the light to a camera summing up to four cameras per module. Each stripe is imaged by the 12×6=72 mirrors. Therefore, it is sliced at 0.375"/12mirrors=0.03125". Each set of 72 mirrors has a corresponding set of pupil mirrors to form the virtual slit and send the light to the camera. The expected throughput in K is ~75%, neglecting the grism.

The current configuration of VESPER includes 2 modules summing to 12 FSs and to 8 cameras, each fed by a 4kx4k detector. The 12 FSs probe an AO-corrected area of approximately 24"×70", as shown in Figure 2. This layout provides a total effective area of 1.7"×1.5"×12, equivalent to 31 arcsec².

## 3 Expected Performances

Figure 6 shows the Encircled Energy (EE) distribution as a function of radial distance at wavelength 2.19 μm (fourth channel) for the central position and four positions offset by ±36" for the NEXUS subsystem. As input, a beam distributed within a Gaussian with FWHM equal to the ELT diffraction limit was considered. The Figure shows that more than 90% of the flux falls within one NEXUS pixel (15 μm, ~35 mas). In Figure 7, the spot diagram obtained for different positions on the field is shown. The Airy disk, the best-focused spot that a perfect lens with a circular aperture can make, falls well within a pixel being the Airy radius ~5.9 μm, almost 1/3 of the pixel size.

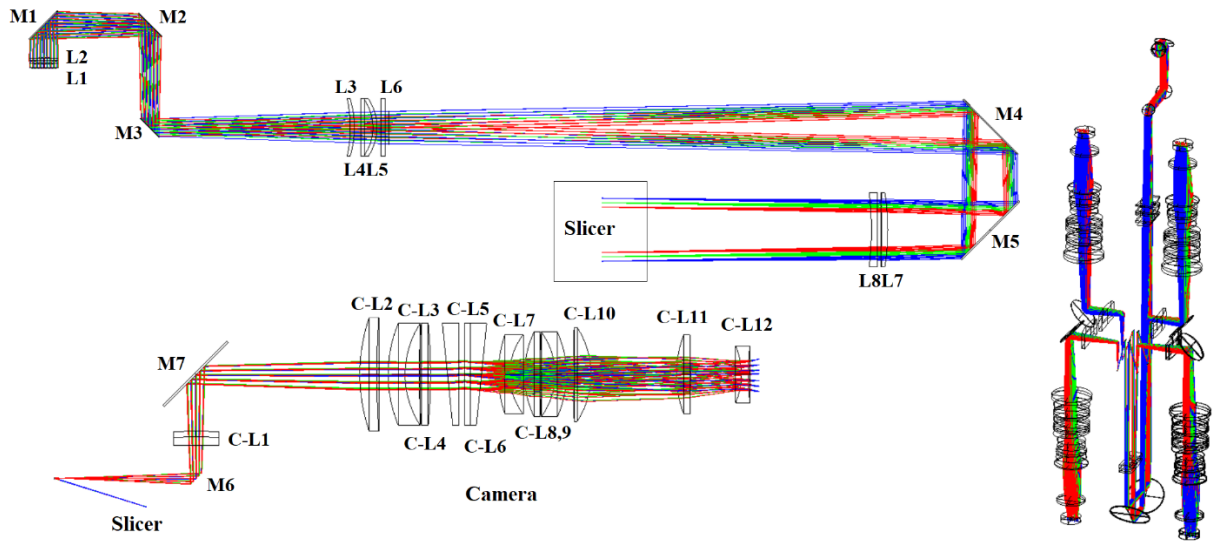

**Figure 4:** VESPER optical design from the focal plane of MORFEO to the four detectors. The integral field selector (IFS), not shown in the figure.

In Figure 8, the encircled energy as a function of the radial distance is shown for a point source falling at three different positions on a slicer mirror. In this case, the diffraction limit of ELT has not been considered. Fig. 9 is similar to Fig. 8, but in this case the point source is modeled as a Gaussian with FWHM equal to the diffraction limit of the ELT. The comparison between the two provides a direct measure of the quality of the VESPER optical design. It can be seen that the image quality is dominated by the optical systems that precede SHARP. In Figure 10, the spot diagram for the three different positions considered in Fig. 9 is shown.

**Table 3:** Overview of the VESPER optical components, including key parameters and design characteristics

| | R1 [mm] | R2 [mm] | Thickness [mm] | Material |
|---|---|---|---|---|
| **The lens from focal panel up to Cameras** | | | | |
| L1 | 392.6 | 181.2 | 10 | SILICA |
| L2 | 196.2 | 532.8 | 10 | CAF2 |
| L3 | 316.5 | 236.8 | 10 | CAF2 |
| L4 | inf | inf | 10 | CAF2 |
| L5 | 122.5 | 128.5 | 10 | SILICA |
| L6 | inf | inf | 10 | SILICA |
| L7 | 588.5 | inf | 15 | CAF2 |
| L8 | inf | 762.7 | 15 | SILICA |
| **Camera** | | | | |
| C-L1 | 11200 | 502.1 | 25 | SILICA |
| C-L2 | 310.9 | 1461.4 | 20 | CAF2 |
| C-L3 | 233.8 | 165.2 | 30 | SILICA |
| C-L4 | 1125.9 | 971.9 | 20 | CAF2 |
| C-L5 | inf | inf | 20,10 | ZNSE, SILICA |
| C-L6 | inf | inf | 10,20 | SILICA, ZNSE |
| C-L7 | 222.9 | 115.8 | 20 | SILICA |
| C-L8 | 174.3 | 1259.4 | 30 | CAF2 |
| C-L9 | 126.4 | 183.04 | 20 | SILICA |
| C-L10 | 13300 | 153.1 | 30 | CAF2 |
| C-L11 | 195.1 | 1797.1 | 25 | BAF2 |
| C-L12 | 125.6 | 404.1 | 12 | SAPPHIRE |

## 4 Mechanical design description

### 4.1 Requirement

The basic mechanical requirements for SHARP are typical of most infrared (IR) cameras. Detectors, optics, and optical bench must be cooled to cryogenic temperatures. The operating temperature of the instrument will be around 70-80 K according to the scientific impacts related to thermal background contributions (The sky brightness in K (at 2.2 mu) is about 13 mag/arcsec^2 (Vega system), corresponding to Isky~10(^-3) W/m^2/mu/sr. The thermal radiation from the instrument reaching the detector can be considered negligible when it is much lower than the limiting accuracy (rms) in the sky estimate. The number of sky photons Nsky is proportional to the exposure time t. The rms, assuming Poisson statistics, is sqrt(Nsky,). Therefore, the accuracy in the sky estimate that can be reached is d=1/sqrt(Nsky,)∝1/sqrt(t), i.e. to double the accuracy the exposure increases 4 times. Exposures longer than 10 hours do not provide significant improvements. For t=10 hr the accuracy is ~0.005%. Therefore, the thermal contribution from the instrument must be significantly lower, e.g. 1%, than this accuracy to be negligible. Assuming the instrument is a black body at temperature T, according to Plank's law, the radiance from the instrument at 2.2 mu must be such that B(2.2mu,T)/Isky<0.01*(5x10-5), that is to say B(2.2mu,T)<10^(-10) W/m^2/mu/sr. Resolving for T, we obtain T<125 K. Actually, this contribution should be integrated over the band width [1.9-2.45] mu, However, for a black body at T~70-80 K, according to the Wien's law, the peak of emission is at ~35-40 mu, extremely far from the band, assuring a fully negligible thermal contribution in this range). Optical mounts must compensate for differential contraction between glass and metal during cooldown, while achieving proper spacing of the optics at cryogenic temperature. Lenses must be centered with respect to the optical axis based on the optical error budget resulting from the optical design. Materials within the cryostat must be vacuum compatible, i.e. low outgassing, and thermal loading must be minimized to increase dewar hold time.

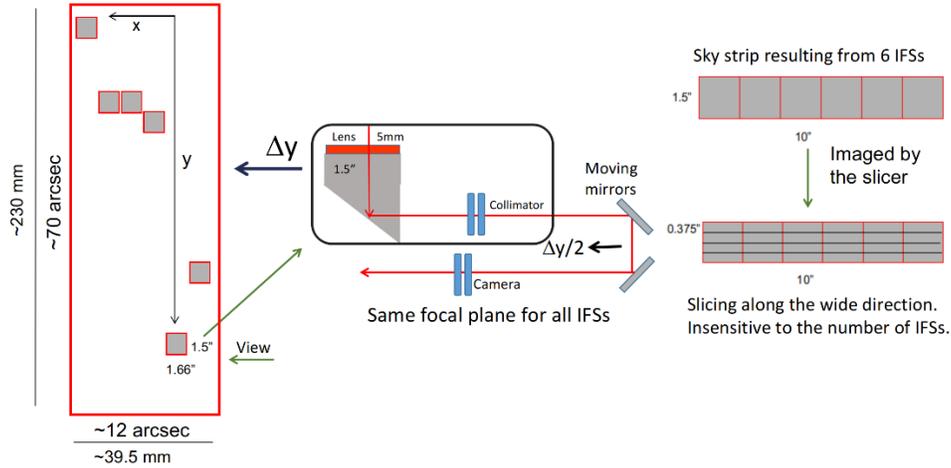

**Figure 5:** Schematic view of the Integral Field Selector system feeding one channel of VESPER. Left - The 6 FSs are aligned along the x direction and separated by 0.33" each other. The small gaps between the FSs are masked on the focal plane to cancel out their contribution to the background due to scattered light. The FSs can be moved along the y direction to sample an area of ≈12"×70". Center - Each FS is composed of a prism having the upper surface with power and a collimator. They rigidly move by Δy. Their movement is compensated by the movement (Δy/2) of the two 45-degree moving mirrors to keep constant the optical path between collimator and camera for all the FSs. Right – The image formed onto the VESPER focal plane by the 6 IFSs is a strip as if they were contiguous. Onto the VESPER focal plane there is the slicer that samples the image in slices of 0.031"×12" (see text for details)

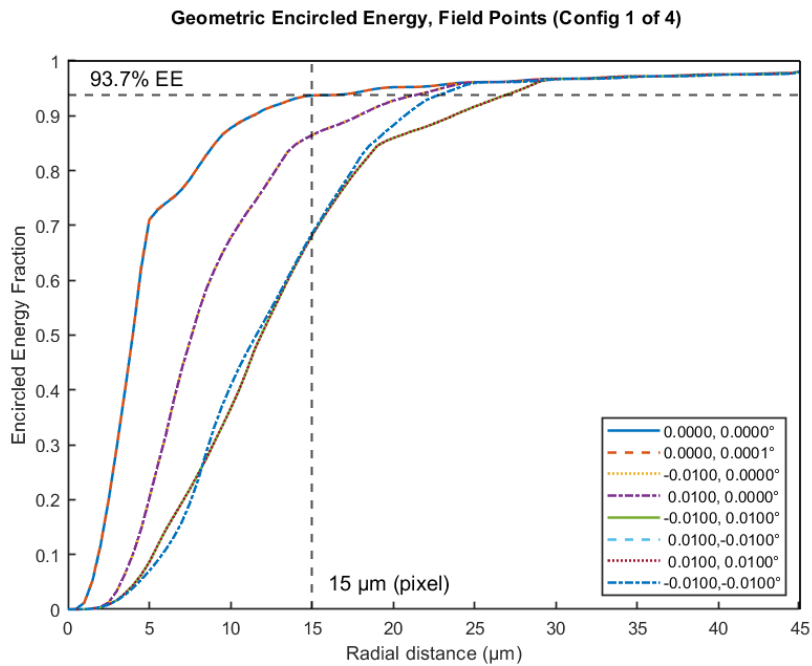

**Figure 6:** Encircled Energy (EE) distribution as a function of radial distance at wavelength 2.19 μm (fourth channel) for the central position and four positions offset by ±36"

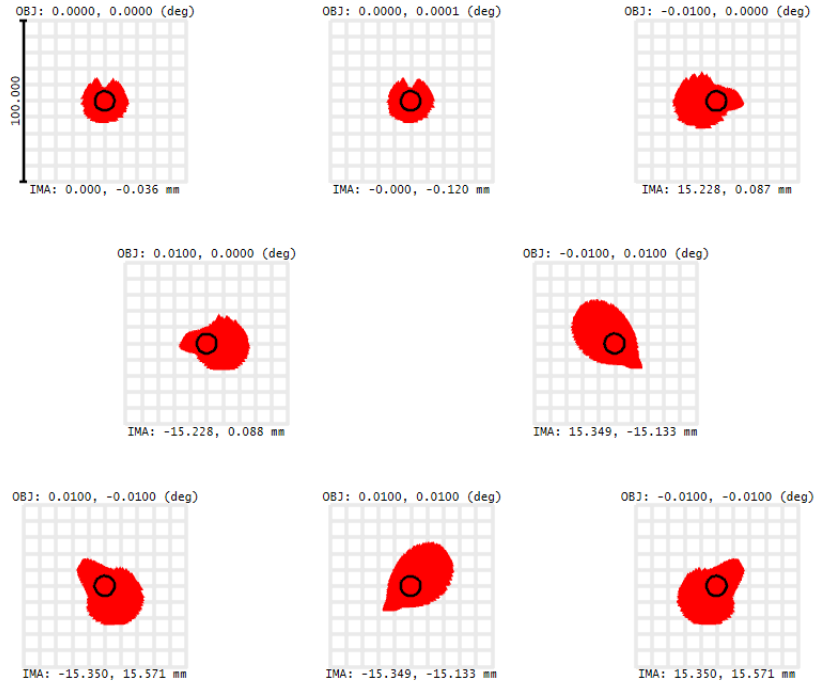

**Figure 7:** Spot diagram for the different positions on the field for the NEXUS subsystem

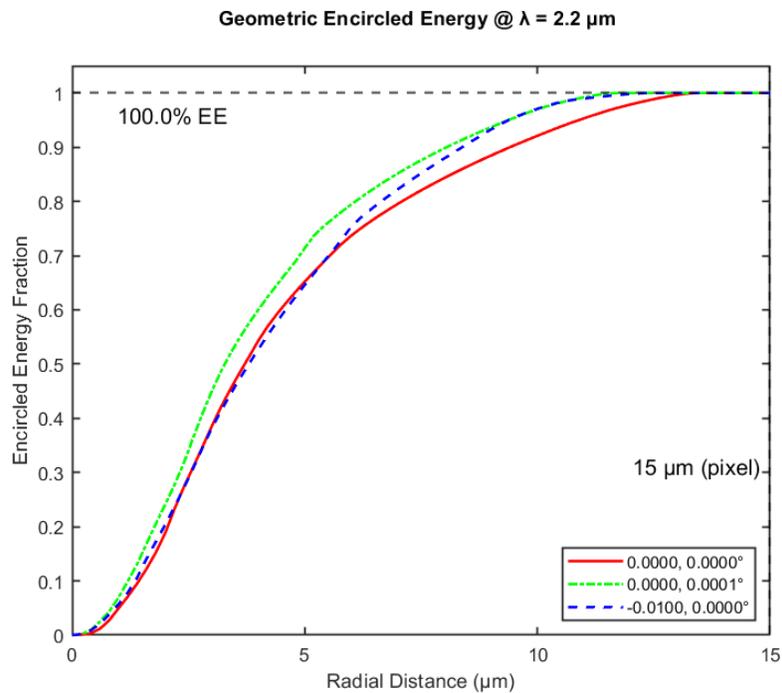

**Figure 8:** Fraction of Encircled Energy (EE) as a function of radial distance for an ideal point source seen by VESPER. The three curves are for three different positions of the source on one of the 288 mirrors of the slicer, namely at the center (red) and at two corners (blue and green).

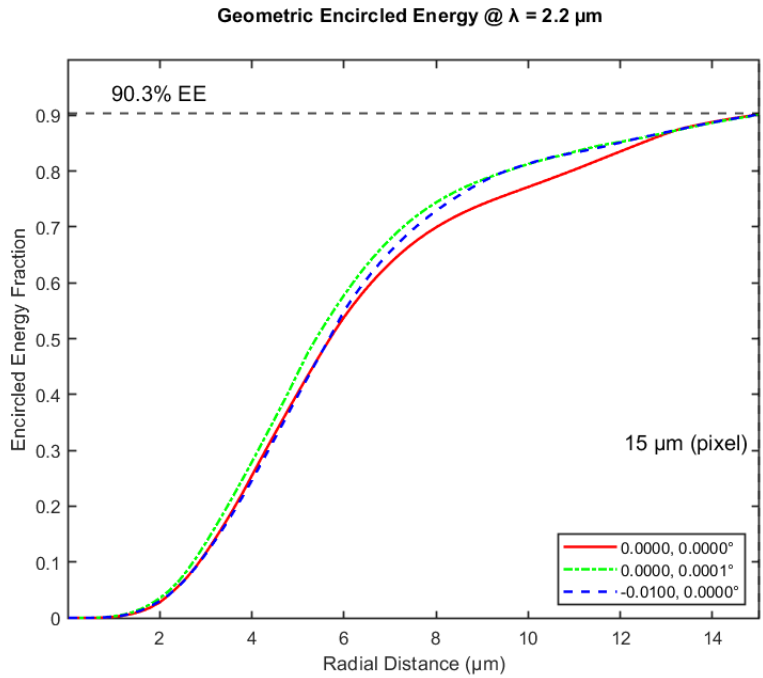

**Figure 9:** Same as Fig. 8 but in this case the point source is modeled considering the diffraction limit of ELT.

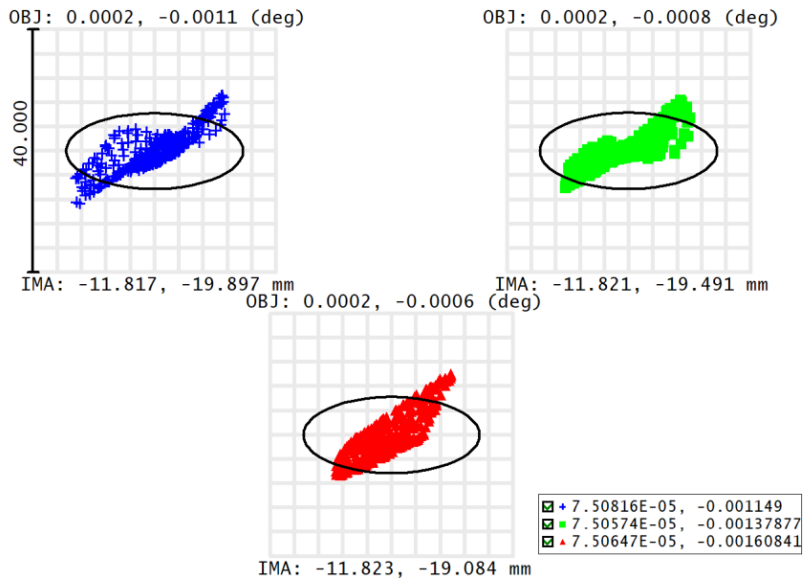

**Figure 10:** Spot diagram measured for the three different positions of Fig. 9.

SHARP also requires complex subsystems: the CSS for NEXUS and the FSS for VESPER. Furthermore, since SHARP utilizes 12 cameras, an intelligent arrangement of the equipment is necessary to minimize the occupied space. The space for installing SHARP on the Nasmyth platform of the ELT including its supporting structure, and all auxiliary equipment (such as cryogenic systems and electrical cabinets) measures approximately 4000 × 5000 × 7000 mm³. Consequently, the target diameter for SHARP is expected to be around 2000–2500 mm. Reducing the overall dimensions not only minimizes material usage but also decreases the power required to maintain cryogenic conditions.

**4.2 Mechanical layout**

The arrangements of the optic elements and equipment are depicted in Figure 11, which are housed within a cryogenic cylindrical tank with dimensions of 2000 mm in diameter and 3000 mm in height. At the entrance of both NEXUS and VESPER, there are two critical subsystems: the CSS for precise mechanical alignment and configuration of the slits, and the FSS for collecting light from specific positions for multi-integral field spectroscopy. The most challenging issue in the initial design phase was arranging these systems compactly, given the configuration of optical entrance mechanisms, filter wheel mechanisms, modules, and cameras. This compact arrangement is advantageous because cooling the tank to cryogenic temperatures is highly time-consuming.

The intent of the design is to make all the mechanisms accessible for servicing after removal of either the front or the rear cap. As the vacuum shell must safely resist the atmospheric pressure load in combination with the loads of external components and also have sufficient stiffness to keep the relative motion between internal components within acceptable limits, a tubular vacuum shell was chosen based on its efficiency in meeting these requirements.

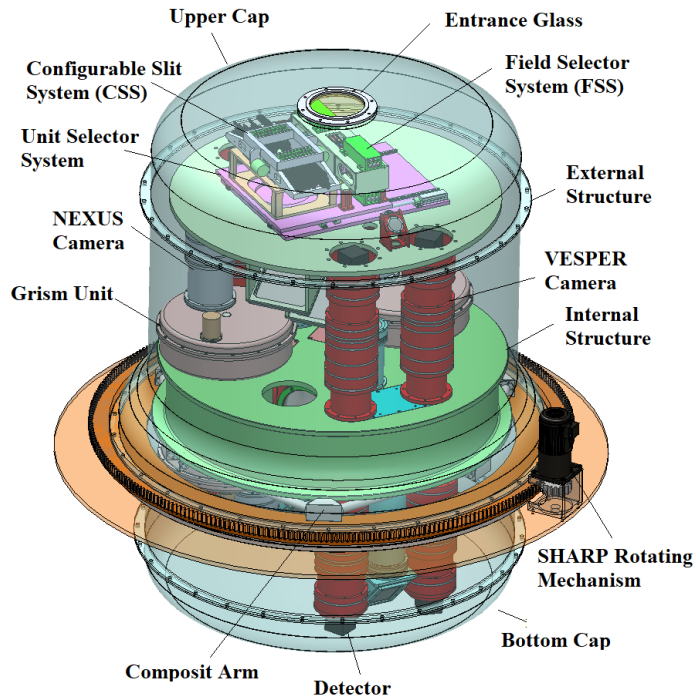

**Figure 11:** SHARP equipment arrangements and conceptual design

To minimize heat transfer through conductivity, all internal equipment are installed in one structure that is only connected with the outer shell through only an insulated bar mechanism (composite arm depicted in Figure 11).

### 4.3 Internal cold structures

The internal structure of SHARP supports all optical elements, including mirrors, collimators, cameras, and filter wheels, as well as two main subsystems. It comprises two primary sections: the middle structure and the upper floor. The middle structure, which is the main part, houses most of the equipment, ensuring stability and precise alignment of the optical components. The upper floor contains the mechanism for switching between the subsystems and accommodates the two main subsystems: The FSS system and the CSS. These two sections are interconnected through the VESPER camera structure, which ensures cohesive operation and structural integrity.

To simplify the assembly process while increasing precision, SHARP is designed modularly. The main structure, as shown in Figure 12, will be manufactured as a single cast part using a 7000-series aluminum alloy, which is suitable for cryogenic conditions. Precise machining will then be applied to both sides of the structure, ensuring accessibility and ease of machining to create accurate reference points for the installation of other components.

The optical elements of NEXUS, which are installed within the main structure, are designed as three separate modules. The first module is depicted in Figure 13a. As shown, this module consists of several smaller components that are precisely manufactured separately, then assembled and calibrated before final installation on the main structure. A 7000-series aluminum alloy is also used for this module and other module components. The other two modules, one of which is shown in Figure 13b, are identical to the first module, each is manufactured, assembled separately, and calibrated before installation.

The concept of the image slicer is illustrated in Figure 14. The input light is initially divided into two strips by Slicer 1. Each of these strips is further divided by Slicers 2 and 3, creating additional sub-strips that are then directed to mirrors 1 through 4. The structure of the image slicer can be manufactured using either welding or casting techniques. After precise machining, it will undergo calibration to ensure accuracy before being installed in the main structure. Finally, the fully assembled main structure is shown in Figure 15.

The components of the upper surface, like those of the main structure, will be assembled and calibrated separately before being installed on the main plate. This plate is manufactured and precisely machined in advance with accurate reference points. The subsystem is shown in Figure 16.

After the final calibration, additional components, including the NEXUS grism wheel, NEXUS cameras, and VESPER cameras, are mounted onto the designated reference points on the assembled main structure. These components are installed either directly onto the main structure or onto the pre-assembled modules, ensuring precise alignment and integration within the system. The final calibration of SHARP is performed between the upper internal structure and the main structure using the existing adjustable folding mirrors positioned between the two modules. The fully assembled internal structure of SHARP is shown in Figure 17.

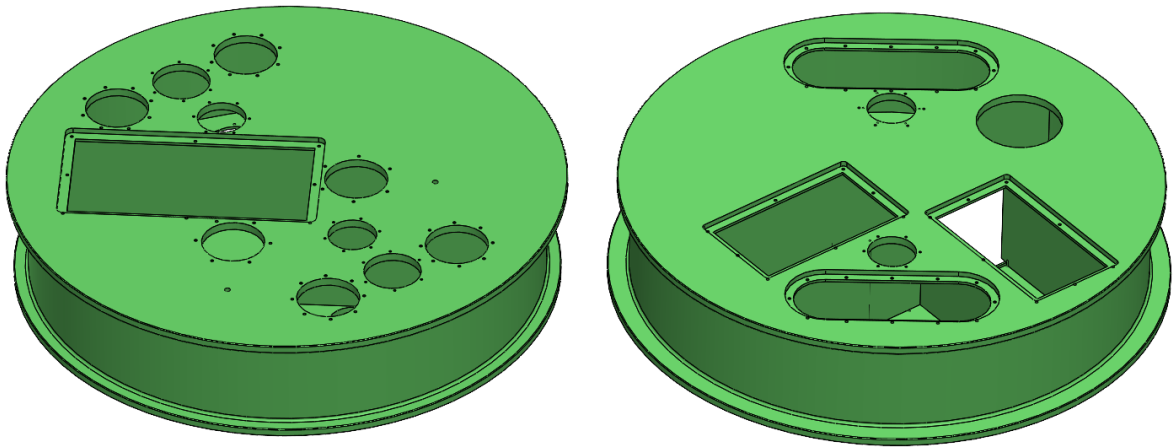

**Figure 12:** SHARP's main structure, cast from 7000-series aluminum and precision-machined for accurate assembly

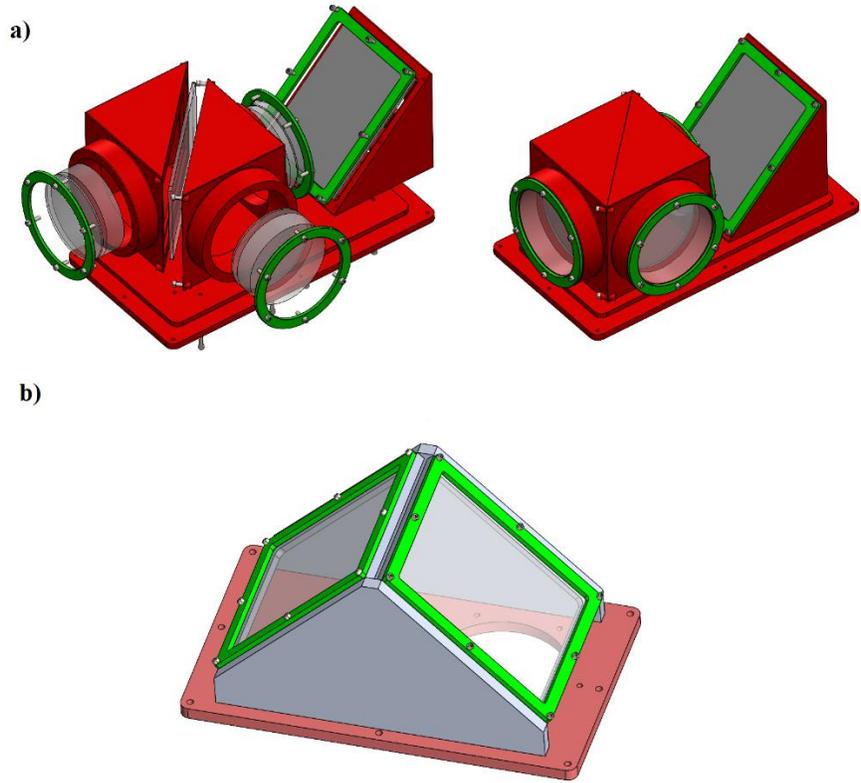

**Figure 13:** The first (a) and second (b) NEXUS optical modules, pre-assembled and calibrated before integration.

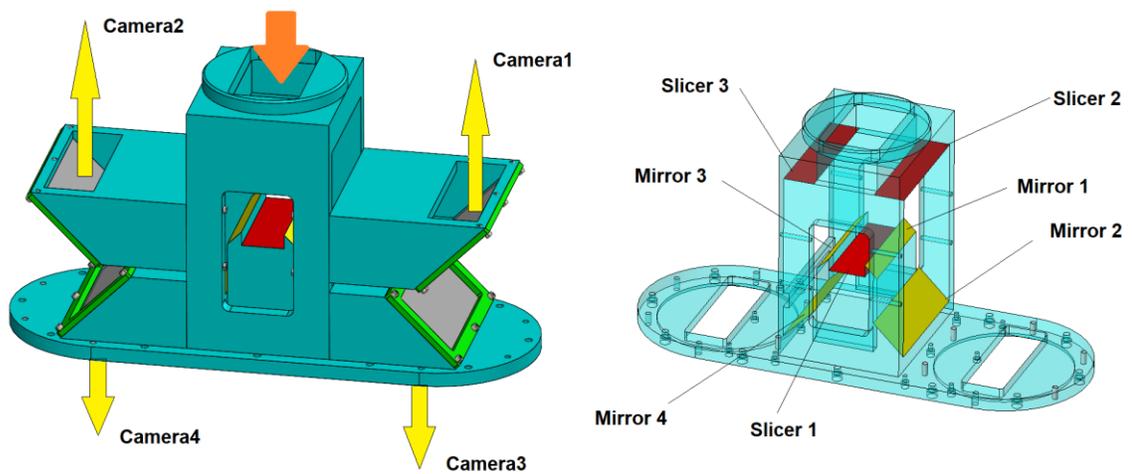

**Figure 14:** Structure of the image Slicer

## 4.4 External structure

Once all the components have been assembled on the internal structure, the entire system is mounted within the cylindrical tank using a specially designed composite arm. This composite arm serves to significantly reduce thermal conductivity between the internal cryogenic environment and the external structure, thus preserving the required low-temperature conditions essential for optimal performance. The external structure is integrated with a high precision thrust bearing and an advanced rotational mechanism. This mechanism facilitates the entire SHARP system's ability to rotate seamlessly around its axis, ensuring accurate and continuous tracking of celestial objects as they move across the sky.

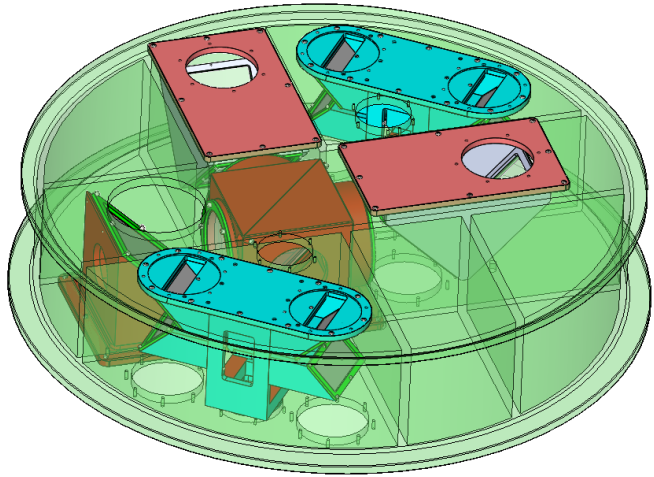

**Figure 15:** final assembly of main structure

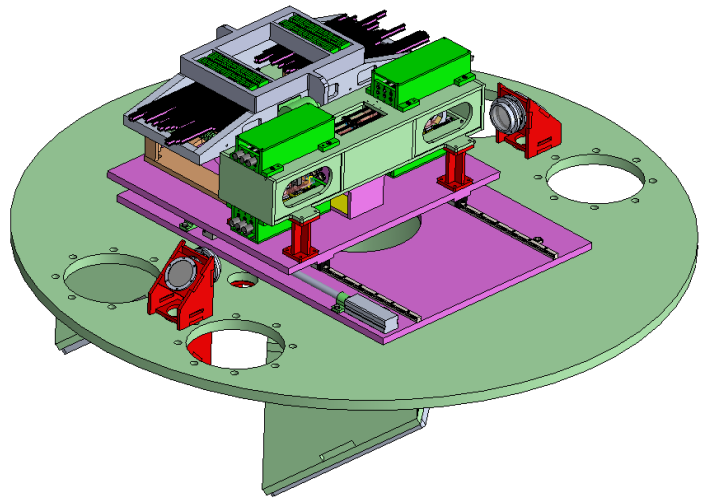

**Figure 16:** Assembled upper surface subsystem of SHARP. The components are individually assembled, calibrated, and mounted onto the main plate, which is precisely machined with reference points for accurate installation.

### 4.5 Unit Selector System

A mechanism is required to switch the optical path between NEXUS and VESPER. The overall concept of this mechanism is illustrated in Figure 18.

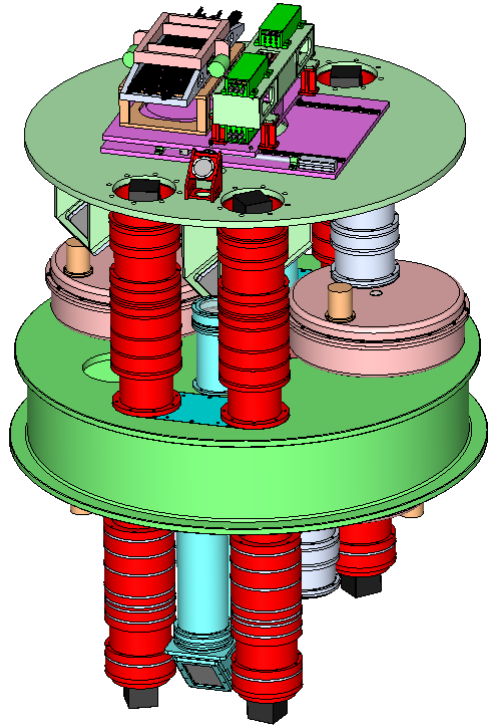

**Figure 17:** Fully assembled internal structure of SHARP.

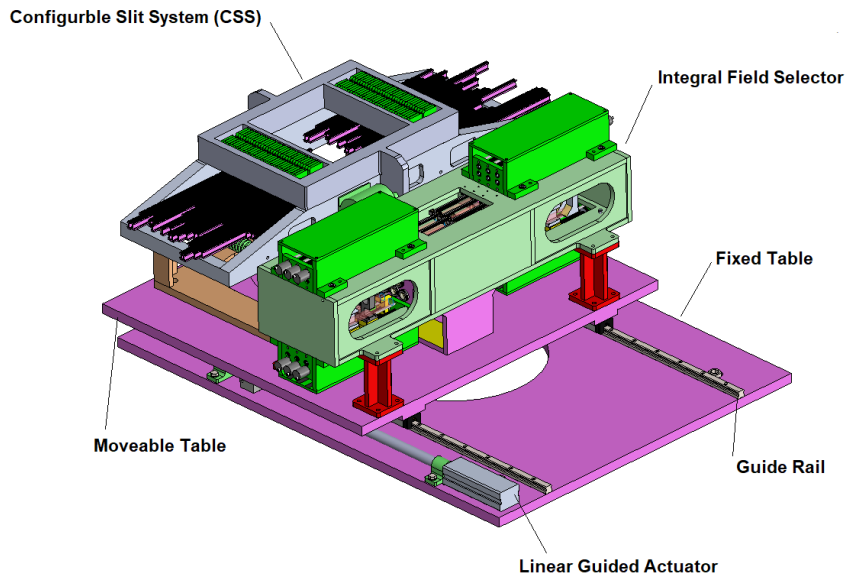

**Figure 18:** Conceptual design of the switching mechanism between NEXUS and VESPER optical paths (Unit Selector System)

It comprises two tables (one fixed and one movable), rails, and a linear guided actuator.

The fixed plate is mounted on the upper floor of the spectrograph, precisely at the light entrance. The two primary mechanisms, the Field Selector System of VESPER and the Configurable Slit

System of NEXUS, are installed on the movable plate and can be moved linearly. By positioning, either mechanism at the center of the spectrograph, light will pass through the selected system. For instance, placing the FSS at the center activates VESPER, while replacing it with the CSS switching the active system to NEXUS.

## 4.6 The Field Selector System (FSS) of VESPER

The conceptual mechanical design of the FSS is depicted in Figure 19 and is crucial for the selection of targets and management of light within the optical system. The FSS employs an array of 12 prisms, which are strategically positioned across two parallel planes.

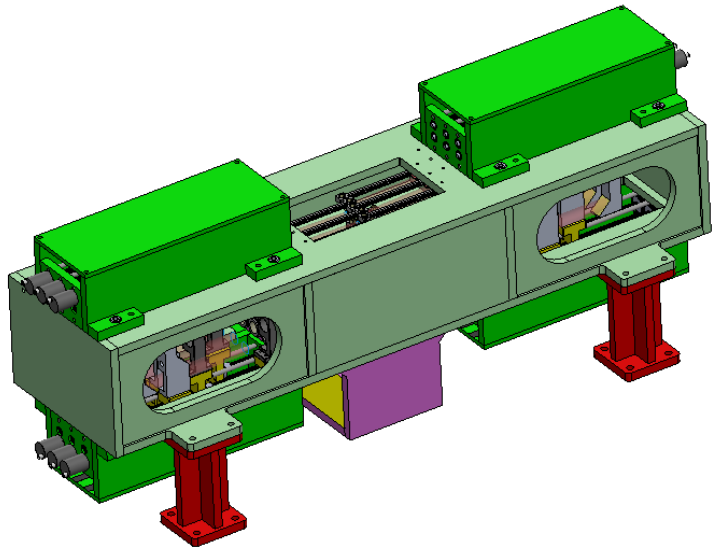

**Figure 19:** Conceptual Design of The Integral Field Selector (IFS) system

In Figure 20, the proposed mechanism for scanning the focal plane of VESPER is illustrated. For the first plane, the collecting prism and its collimator are mounted on a rigid frame. This frame is connected to a sturdy guide rail that enables movement in the y direction, powered by a stepper motor and a ball screw mechanism, as shown in the figure.

To maintain constant the optical path between the collimator and the camera, two folding mirrors are incorporated into the design. The movement of these folding mirrors is also driven by a ball screw mechanism. However, the same stepper motor that drives the collecting prism is utilized to power the folding mirrors, with the motion transmitted through a gear pair. Since the folding mirrors are required to move at half the displacement of the collecting prism, a gear ratio of 0.5 is implemented. Alternatively, the movement of the folding mirrors can be achieved using an independent stepper motor, similar to the one used for collecting prism.

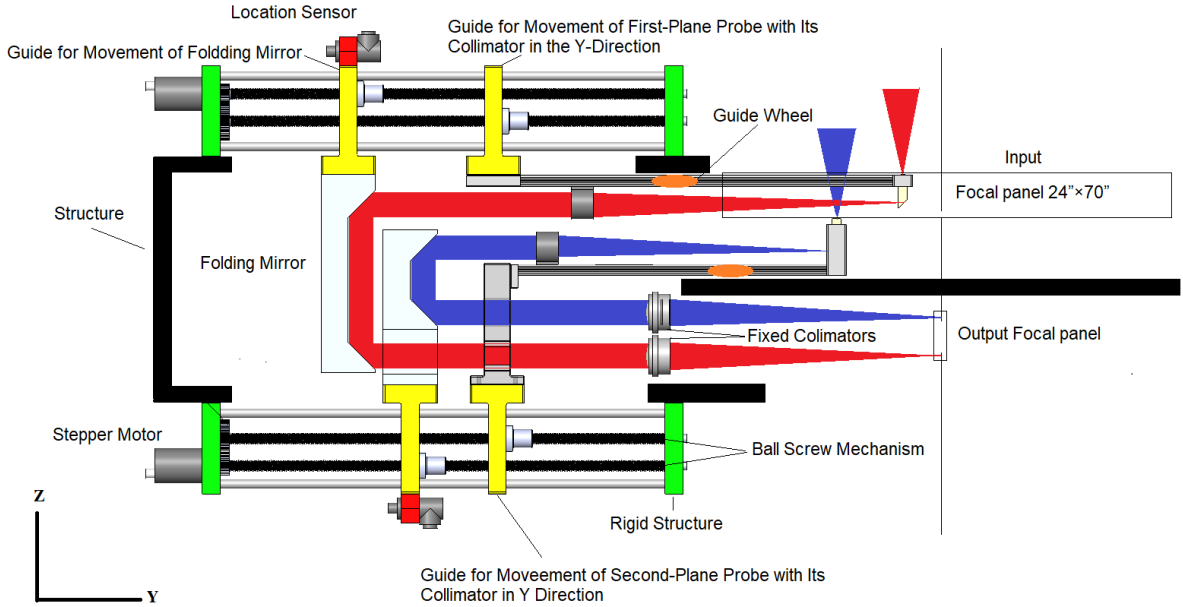

**Figure 20:** Proposed scanning mechanism for the VESPER focal panel.

While the stepper motor is equipped with a position sensor to enable closed-loop position control, an additional location sensor is installed on the guide rail for the folding mirrors to ensure precise control. For enhanced control and redundancy, a second position sensor could be added to the guide rail for the collecting prism. In addition to the mentioned sensors, the quality of the captured images can also be integrated into the control system.

For the second plane, the same design concept is applied, with the key difference being that the guiding mechanism is installed at the bottom of the structure. Based on the design geometry, the FSS structure prevents light from penetrating the focal plane through the gaps between the selector prisms.

The selected light from the 12 prisms is then directed to two VESPER modules (6 prisms per module), located on either side of the FSS system. This configuration is designed to prevent interference and vignetting by carefully managing the spatial arrangement of the optical components. The detailed opto-mechanical design of this system will be presented in a forthcoming dedicated paper.

### 4.7 CSS of NEXUS

The CSS of NEXUS is conceptually similar to the one described in Henine et al[13] and Spanoudakis et al[15] feeding both the MOSFIRE and EMIR[9-11] instruments at the Keck[14] and GTC observatories respectively. As shown in Figure 21 and 22, a slit is formed at the position of the selected object by translating two opposite bars toward each other on the focal plane of NEXUS. The two bars block the light coming from outside the slit formed by the bars themselves. Both slit position and slit width are adjustable and are controlled with a micrometric precision[13]. According to the recommendations of the SHARP Science Team, the minimum length of the configurable slit is ~2.4", assuring a good sky sampling also for extended sources, namely for high-redshift galaxies. This limits the maximum number of slits to 30. Longer slits can be obtained by aligning two or more slits, resulting in a slit length multiple of 2.4". Slits are all on one orientation. We are studying a system to feed the slits with an inversion prism to rotate by an arbitrary angle the field subtended by the slits (see Saracco et al[8].).

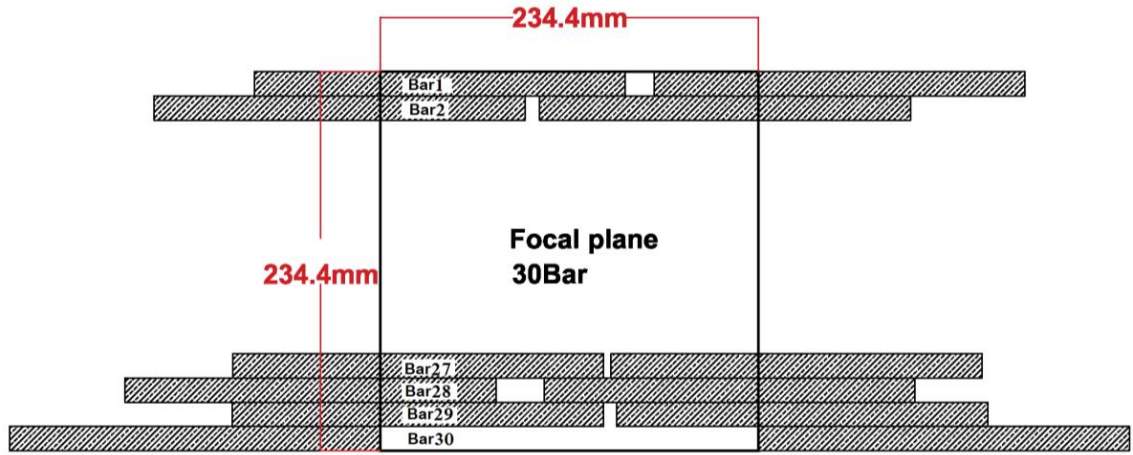

**Figure 21:** Conceptual view of the masking mechanism. Pairs of bars move from each side of the focal plane to form a slit and mask the targets. When the bars move back toward their fully open position, the focal plane is unobstructed.

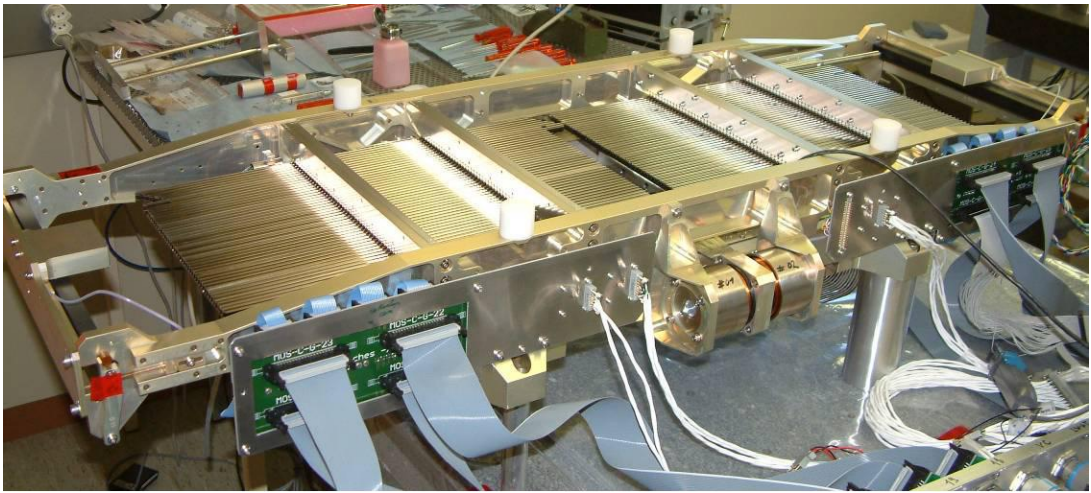

**Figure 22:** The configurable slit-mask unit for the Keck telescope designed and manufactured by CSEM co.

## 4.8 Grism wheel mechanism

The Grisms Unit (GU) is a critical component of the optical system, designed to accommodate up to three grisms along with one open position to allow pre-imaging for field recognition, all mounted on a precision-engineered rotating wheel with an external diameter of 670 mm. As depicted in Figure 23, the GU assembly comprises several key elements, including the housing cover, the rotating wheel, the drive unit, and the grisms securely mounted in their respective mechanical holders. This intricate setup allows for the seamless interchange of optical elements within the spectrograph.

The rotational movement of the grism wheel is actuated by a cryogenic stepper motor, chosen for its reliability and precise control in low-temperature environments. The motor drives an internal gear that is integrally connected to the grism wheel, facilitating smooth and accurate positioning of each grism. The cover housing supports the drive motor, the grism wheel, and the NEXUS camera, all while maintaining thermal stability within the cryogenic conditions.

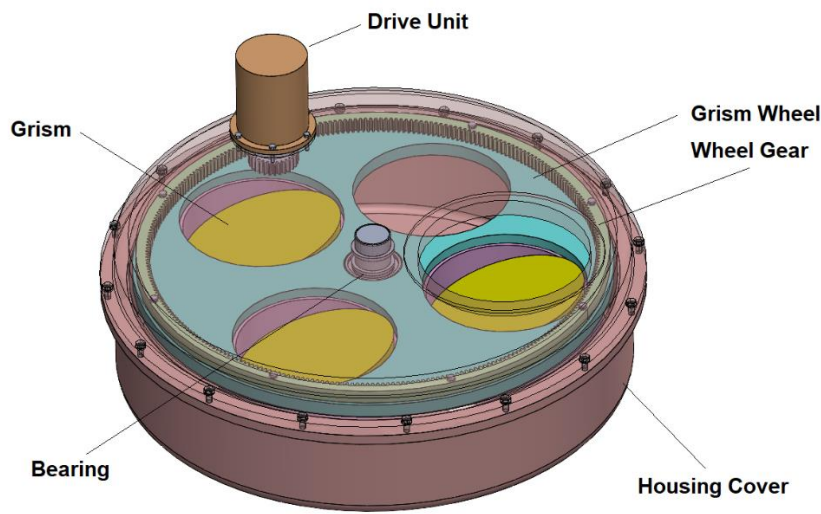

**Figure 23:** Grism wheel mechanism

To ensure the mechanical integrity and precision of the system, a single angular contact bearing is employed as a radial constrainer. This bearing works in tandem with a friction plain bearing, which serves a dual purpose: it constrains the axial and tilt displacements of the wheel and generates the necessary friction torque to prevent backlash. The friction torque also provides a braking force when the motor is depowered, ensuring the stability of the optical components during operation and minimizing any potential for misalignment.

It is worth noting that in this mechanical design, the grism wheels are independent of each other. Therefore, each camera can in principle provide a different spectral resolution according to the user's needs.

## 5    Conclusions

The presented work primarily outlines the conceptual design of SHARP, encompassing both the overall system architecture and its key subsystems. While this paper provides a general reference framework, it is the beginning of a more extensive development process. Detailed designs of the full system and its subsystems are planned for future work. In that phase, we will conduct thorough Finite element Analysis (FEA) to evaluate how the design reacts under nearly real conditions. Those analyses will cover thermal performance, structural integrity, and optical error assessments. The insights gained will be used to refine and optimize the spectrograph's performance. Moreover, the subsequent phases will include the fabrication of prototypes for the SHARP systems or for individual subsystems of SHARP. These prototypes will be instrumental in validating the design through testing, ensuring that SHARP meets the demanding scientific and technical requirements. By progressing from conceptual design to detailed analysis and prototyping, SHARP aims to set a new benchmark in near-IR spectroscopy, leveraging the unprecedented capabilities of the next generation ground-based ELTs and space-based telescopes to explore the Universe with unparalleled depth and precision.

**Disclosures**

The authors declare there are no financial interests, commercial affiliations, or other potential conflicts of interest that have influenced the objectivity of this research or the writing of this paper

**Code and Data Availability**

All data in support of the findings of this paper are available within the article. More information can be found in this website: https://sharp.brera.inaf.it/

**Acknowledgements**

HM and the SHARP team acknowledge support by grant "Bando INAF Ricerca Fondamentale 2022", Techno-Grant SHARP - 1.05.12.02.01

**Hossein Mahmoodzadeh** is a Mechanical Engineer with over 12 years of industry experience and has been a key member of design teams in several research and industrial projects. He earned his M.S. degree from the University of Tabriz and is currently pursuing a Ph.D. at Politecnico di Milano. Concurrently, he works as a Mechanical Expert and Researcher at INAF – Osservatorio Astronomico di Brera in Milan, Italy.

**Paolo Saracco** is Prime Researcher at INAF - Osservatorio Astronomico di Brera in Milan, Italy. He received his MS degrees in Astronomy from the University of Bologna in 1992 and his PhD degree in Astrophysics from the University of Milano 1995. He is the author of more than 85 refereed journal papers on galaxy evolution. His current research interests include instrumentation for the ELT